\documentclass[11pt,onecolumn,a4paper]{IEEEtran}
\usepackage{rotating}
\usepackage[bottom=3cm]{geometry}
\usepackage{amsthm}%
\usepackage{parskip}
\makeatletter
\def\thm@space@setup{%
   \thm@preskip=\parskip \thm@postskip=0pt
}
\makeatother

\usepackage{cite}
\usepackage{amsmath,amssymb,amsfonts}
\usepackage{graphicx}
\usepackage{textcomp,nicefrac}
\def\BibTeX{{\rm B\kern-.05em{\sc i\kern-.025em b}\kern-.08em
T\kern-.1667em\lower.7ex\hbox{E}\kern-.125emX}}
\newif\ifTwoColumn
\usepackage{amsmath,amssymb,amsfonts,amsthm}
\usepackage{mathtools}
\usepackage{array}
\usepackage{mymath}
\usepackage{nth}
\usepackage{caption}
\usepackage{todonotes}
\usepackage{arydshln} 
\usepackage{MnSymbol} 
\usepackage{overpic}
\usepackage{varwidth, tikz}
\usetikzlibrary{shapes, arrows, shapes.misc, arrows.meta, positioning, matrix, calc, fit, fadings, patterns}
\usetikzlibrary{calc,patterns,decorations.pathmorphing,decorations.markings,arrows, arrows.meta}
\usepackage{placeins} 
\usepackage{textcomp,nicefrac}
\def\BibTeX{{\rm B\kern-.05em{\sc i\kern-.025em b}\kern-.08em
T\kern-.1667em\lower.7ex\hbox{E}\kern-.125emX}}

\newlength{\lw}
\newcommand{\sst}[1]{\footnotesize{#1}}

\newtheorem{lemma}{Lemma}
\newtheorem{definition}{Definition}

\newtheorem{proposition}{Proposition}

\newcommand\Tstrut{\rule{0pt}{2.6ex}}       
\newcommand\Bstrut{\rule[-0.9ex]{0pt}{0pt}} 
\newcommand\TLstrut{\rule{0pt}{3.6ex}}       
\newcommand\BLstrut{\rule[-1.8ex]{0pt}{0pt}} 
\newcommand{\TBstrut}{\Tstrut\Bstrut} 
\newcommand{\TBLstrut}{\TLstrut\BLstrut} 

\date{\today}

\begin{document}
\title{Symmetry Exploitation in Orbit Feedback Systems of Synchrotron Storage Rings}
\author{Idris Kempf, Paul J. Goulart, Stephen R. Duncan, and Guenther Rehm
\thanks{The research leading to these results is supported by Diamond Light Source and the Engineering and Physical Sciences Research Council (EPSRC) with a CASE studentship.}
\thanks{I. Kempf, P. Goulart and S. Duncan are with the Department of Engineering Science, University of Oxford, Parks Road, Oxford, OX1 3PJ, UK; (e-mail: \{idris.kempf,paul.goulart,stephen.duncan\} @eng.ox.ac.uk).}
\thanks{G. Rehm was with Diamond Light Source, Didcot, OX11 0DE, UK; He is now with BESSY, Berlin, 12489, Germany (e-mail: guenther.rehm@helmholtz-berlin.de).
}
}

\maketitle
\begin{abstract}
Structural symmetries in the storage ring of synchrotrons are intentionally created during the design phase of the magnetic lattices, but they are not considered in the design of control algorithms that stabilize the beam of accelerated particles. The choice of control algorithm, however, is limited by the speed requirements of the synchrotron. Standard control algorithms for synchrotrons are based on a singular value decomposition (SVD) of the orbit response matrix. SVD controllers neither exploit the structural symmetries nor exhibit any speed advantages. Based on the periodicity and the reflection properties of the betatron function, we show that these structural symmetries are inherited by the orbit response matrix. We show that the resulting block-circulant and centrosymmetric properties of the matrix can be used for different computationally efficient decompositions of the controller. We also address the case of broken symmetry due to odd placements of magnets and monitors. Our efficient decomposition could enable the use of more advanced control techniques for synchrotrons, such as control algorithms that require real-time optimization. These advanced control techniques could in turn increase the quality of research in synchrotron light sources.
\end{abstract}

\begin{IEEEkeywords}
Orbit Feedback, Synchrotron, Symmetries
\end{IEEEkeywords}

\section{Introduction}
\IEEEPARstart{I}{n} most synchrotrons the magnetic lattices, the beam position monitors and the corrector magnets are placed in repeated patterns around the storage ring~\cite[Ch. 10.2.4, p. 329]{WIEDEMANN}. These repeated sections are usually referred to as \emph{superperiods} or \emph{cells}, and their pattern invokes a circulant symmetry. Often an additional symmetry is introduced by mirror-reflecting the pattern in the middle of the storage ring, which invokes a centrosymmetry. The circulant pattern considerably simplifies the design of the synchrotron, while the mirror-reflection cancels out non-linear effects introduced by quadrupole and sextupole magnets. Although this symmetry is intentionally created in the design phase of the synchrotron, it is most often not considered during the synthesis of the orbit feedback system. The feedback system is designed around the \textit{orbit response matrix} $\mathbf{R}\in\R^{SN_B\times SN_C}$, where $S$ is the number of cells and $N_B$ and $N_C$ the number of monitors and corrector magnets per cell, respectively, that relates the corrector magnet inputs $\mathbf{u}(\inv{z})$ to the horizontal (or vertical) trajectory error of the electron beam $\mathbf{y}(\inv{z})$ as
\begin{align}\label{eq:ormeq}
\mathbf{y}(\inv{z}) = \mathbf{R}g(\inv{z})\mathbf{u}(\inv{z})+\mathbf{d}(\inv{z}),
\end{align}
where the scalar transfer function $g(\inv{z})$ represents the temporal dynamics of the corrector magnets, which are assumed to be identical for all magnets, $\mathbf{d}(\inv{z})$ the disturbances acting on the electron beam, such as vibrations transmitted through the girders, and $z$ the $\mathcal{Z}$-transform variable, respectively. The element on row $m$ and column $n$, $\mathbf{R}_{m,n}$, of $\mathbf{R}$ is characterized by the  \emph{betatron function} $\beta:\R\rightarrow\R_+$, where $L$ denotes the circumference of the orbit, and given by~\cite[eq.~(2)]{YUDFTPA}
\begin{align}\label{eq:ormelem}
\mathbf{R}_{m,n}= \frac{\sqrt{\beta_m^B\beta_n^C}}{2\sin(\pi Q)}\cos\left(\pi Q-\abs{\phi_m^B-\phi_n^C}\right),
\end{align}
where $\beta_k^X\eqdef\beta(s_k^X),\phi_k^X\eqdef\phi(s_k^X)$ with $s$ representing distance along the storage ring (starting from an arbitrary reference point) and $B$ and $C$ refer to beam position monitors and corrector magnets, respectively. The \textit{phase advance} $\phi:\R\rightarrow\R_+$ is defined as
\begin{align}\label{eq:phaseadvance}
\phi(s_x) \eqdef \int_{0}^{s_x} \inv{\beta}(s)ds,
\end{align}
and, for a stable electron beam, the \textit{betatron tune} $Q\eqdef\phi(L)/2\pi$ is always a fractional number~\cite{MARTINIPA}. The matrix $\mathbf{R}$ typically has a few hundred rows (monitors) and a few hundred columns (corrector magnets) and the feedback loop is operated at a frequency of $10$-$100$ kHz. The magnetic lattices guide and confine the electron beam around the ring, while the orbit feedback system reduces the trajectory error to a few micrometers. The trajectory error must be minimized in order to retain certain properties of the synchrotron light that is used in the beamlines, which are end-stations that use the light for various kinds of experiments, particularly in the X-ray region of the electromagnetic spectrum~\cite{SYNCRAD}.

Most orbit feedback systems use a singular value decomposition (SVD) of $\mathbf{R}$ to synthesize an orbit controller, such as in~\cite{PLOUVIEZ} for the European Synchrotron Radiation Facility, \cite{SANDIRACONTROLDESIGN} for Diamond Light Source, \cite{MAXIVORBIT} for the MAX IV synchrotron or \cite{APSCOMMISSIONING} for the Advanced Photon Source. In these case, the control input is calculated from $\mathbf{u}(\inv{z})=-\mathbf{K}c(\inv{z})\mathbf{y}(\inv{z})$, where $c(\inv{z})$ is a scalar transfer function that accounts for the dynamics $g(\inv{z})$, and the gain matrix $\mathbf{K}$ is calculated from~\cite[eq.~(5)]{SANDIRAOPTIMAL}
\begin{align}\label{eq:ctrmat}
\mathbf{K}\!=\!V_1\inv{(\Sigma^2\!+\!\mu\I)}\Sigma\,U^*\!,
\text{ with }
\mathbf{R}=U\left[\Sigma\,\,\,0\right]\begin{bmatrix}
V_1^*\\V_2^*
\end{bmatrix},
\end{align}
where the zeros in the SVD account for the case that there are usually more actuators than monitors, i.e. $SN_B\!<\!SN_C$. The gain matrix $\mathbf{K}$ is a pseudo-inverse of $\mathbf{R}$ with a regularization parameter $\mu\in\R_+$ that accounts for the large difference between the minimum and maximum non-zero singular value of $\mathbf{R}$, which is common to orbit response matrices of synchrotrons~\cite{PAROBUST}. The SVD approach is applicable to any kind of system but does not exploit the advantages provided by the symmetric structure. Symmetry is always accompanied by certain redundancies in the mathematical representation of the system~\cite{ANDREADISTRIBUTED}, and considering the symmetry speeds-up the controller computations and reduces the memory requirements~\cite{BCINVADMMCONF}. It also benefits the parameter identification~\cite{CIRCSYSID} and the modeling of parameter uncertainty~\cite{PAROBUST}.

There may be different reasons that feedback systems have not generally exploited these existing symmetries (one exception is~\cite{SAJJADPADFT}). Firstly, the matrices involved in the feedback systems have around 100,000 elements and recognizing these symmetries by by inspection is not straightforward. Secondly, the symmetry is often broken by space constraints or by singular components in the storage ring, such as the injection device. Although for this case an SVD seems to be advantageous because it is not reliant on symmetry requirements, it introduces additional difficulties when the controller robustness is verified with respect to parameter uncertainty~\cite{PAROBUST}. Finally, symmetric decompositions are more difficult to find and in contrast to SVDs, which are supported by most numerical software, no widespread implementation of an algorithm for symmetric decompositions exists; symmetric decompositions are often found through prior knowledge of the system structure~\cite{ANDREADISTRIBUTED}.

This paper aims to close the gap between the design of the synchrotron, which intentionally introduces structural symmetries, and the orbit feedback system, which uses the SVD and therefore ignores the structural symmetries. We show that the orbit response matrix inherits the circulant symmetry and/or the centrosymmetry. We present the block-circulant, the centrosymmetric and two different combined decompositions that are possible when both structural symmetries are present. We  illustrate the decompositions using the Diamond-II orbit response matrix. We also address the unavoidable case of broken symmetry. For each of the block-circulant, centrosymmetric and the combined-symmetry cases, we derive formulae for approximating the orbit response matrix using a matrix that has the symmetric properties and minimizes the Frobenius norm error. We show how the asymmetry of the orbit response matrix can be concentrated in certain elements of the symmetrical decomposition. We conclude our analysis by demonstrating the main advantage of exploiting structural symmetries: increased computation-speed. Using a C-language implementation on the device used for the real-time orbit feedback of the Diamond-II and the APS synchrotrons, we compare the SVD approach with the different symmetrical decompositions and demonstrate the improvement in computation speed. We show that accelerating the controller computations reduces the time-delay and allows for faster sampling rates or for the deployment of more advanced control algorithms, such as algorithms that use real-time optimization~\cite{BCINVADMMCONF}.

This paper is structured as follows. In Section~\ref{sec:prelim}, block-circulant and centrosymmetric matrices are briefly outlined. More details are included in Appendix~\ref{app:matrices}. Section~\ref{sec:main} presents our results on the symmetric structure of the orbit response matrix and its decompositions and Section~\ref{sec:brokensymmetry} addresses the case of broken symmetry. In Section~\ref{sec:casestudy}, our results are summarized in a case study of the Diamond-II synchrotron, in which the controller is simulated using symmetric approximations, the nominal stability is verified and the speed advantages of a controller that exploits the symmetric structures are demonstrated.

\section{Preliminary Technical Material\label{sec:prelim}}
\subsection{Notation}
The set of real, strictly positive real and complex numbers are denoted by $\R$, $\R_+$ and $\C$, respectively, with $\sqrt{-1}=i$. For matrices $A$ and $B$, let $A\otimes B$ denote the Kronecker product, $\diag\left( A,B\right)$ the block diagonal concatenation and $A\circ B$ the Hadamard (elementwise) product of two matrices. Let $\I_n$ represent the identity matrix in $\R^{n\times n}$.  For a scalar, vector or matrix $a$, let $a^*$ denote its Hermitian transpose and $\bar{a}$ its element-wise complex conjugate. Let $\Real{a}$ and $\Imag{a}$ denote its real and imaginary part, respectively. For $A\in\R^{m\times n}$, let $\frobnorm{A}$ denote its Frobenius norm that is defined as
$
\frobnorm{A}=\sqrt{\sum_{i=1}^m\sum_{j=1}^n a_{ij}^2},
$
where $a_{ij}$ denotes the (scalar) entry of $A$ at row $i$ and column $j$. Let $\sgn$ represent the signum function and $\modu$ the modulo operator.

\subsection{Matrices with Symmetric Structures}
In an orbit feedback system, the calculation of optimal set-points for the corrector magnets requires at each sampling instant a matrix-vector multiplication. It will be shown that the matrix inherits certain symmetry properties from the storage ring and that these properties can be used to simplify the computations needed for the orbit feedback system.
\begin{definition}\label{def:bc}
Let $\BC(n,p,m)\subset\R^{np\times nm}$ denote the set of \emph{block-circulant matrices} of order $n$ that have the form
\begin{align}
B=\CircBB,\quad\inR{b_i}{p}{m}.\label{eq:bc}
\end{align}
\end{definition}
\noindent Consider the Fourier matrix $\inC{F_n}{n}{n}$, defined as
\begin{align}\label{eq:fouriermatrix}
F_n=\frac{1}{\sqrt{n}}
\begin{bmatrix}
1 & 1 & \dots & 1\\
1 & w & \dots & w^{n-1}\\
\vdots & \vdots & &\vdots\\
1 & w^{n-1} & \dots & w^{(n-1)(n-1)}
\end{bmatrix},
\end{align}
with $w=e^{i\frac{2\pi}{n}}$ and $F^* F = \I_n$. Every block-circulant ($\BC$) matrix is block-diagonalized by the Fourier matrix~\cite{CIRCBOOK},
\begin{align}
\hat{B}=\left(F_n^*\otimes\I_p\right) B \left(F_n\otimes\I_m\right) = \diag\left(\nu_0,\dots,\nu_{n-1}\right),\label{eq:bdiagB}
\end{align}
with $\inC{\nu_i}{p}{m}$ and $\left(F_n^*\otimes\I_p\right)\left(F_n\otimes\I_p\right)=\I_{np}$. Equivalently, the block $\nu_j$ can also be obtained from
\begin{align}\label{eq:diagblocksB}
\nu_j = \sum_{k=0}^{n-1} b_k e^{-i\frac{2\pi jk}{n}}.
\end{align}
The product $F_nx$ yields the coefficients of the discrete Fourier transformation of the vector $x$. Because the Fourier matrix appears in \eqref{eq:bdiagB}, the computation speed of a matrix-vector multiplication $Bx$ can be increased significantly by transforming it to the Fourier domain, i.e. by computing $Bx= \left(F_n\otimes\I_p\right) \hat{B}\left(F_n^*\otimes \I_m\right) x$. The computational efficiency arises from the possibility to employ $m$ parallel Fast Fourier Transformations for computing products like $\left(F_n^*\otimes \I_m\right) x$ and the fact that $\hat{B}$ is block-diagonal. For the case that all elements of $B$ are non-zero, the computation time is reduced by a factor\footnote{This formula serves as a rough estimate of the reduction in computation time and does not consider the details of the implementation, such as the complex arithmetic and the structure of the block-diagonalized matrix.}
\begin{align}
\left(npm + (p+m)n \log_2 n \right)/\left(n^2pm\right).\label{eqa:speedup}
\end{align}
\begin{definition}\label{def:cs}
Let $\mathcal{CS}(q,t)\subset\R^{2q\times 2t}$ denote the set of \emph{centrosymmetric matrices} of the form
\begin{align}\label{eq:rs}
R = \left[\begin{array}{c;{4pt/3pt}c}
r_{1} & r_{2}\\[0.3em]
\hdashline[4pt/3pt]\\[-1em]
J_q r_{2} J_t & J_q r_{1} J_t
\end{array}\right],\quad\inR{r_i}{p}{t},
\end{align}
where $J_k={\arraycolsep=2pt%
\left[
  \begin{array}{@{}*3c}
 & & 1\\[-3pt] & \udots &\\[-3pt] 1 & &
  \end{array}
\right]}\in\R^{n\times n}$ and with $RJ_{2t}=J_{2q}R$.
\end{definition}
\noindent A centrosymmetric ($\CS$) matrix is block-diagonalized by~\cite{CANTONI}
\begin{align}\label{eq:bdiagrefl}
\tilde{R}=\trans{T}_q R T_t =
\diag\left(r_1-r_2 J_t,r_1+r_2 J_t\right),
\end{align}
where the centrosymmetric transformation is defined as
\begin{align}\label{eq:centromatrix}
T_k = \frac{1}{\sqrt{2}}
\begin{bmatrix}
\I_k & \I_k\\
-J_k & J_k
\end{bmatrix}\in\R^{2k \times 2k},
\end{align}
with $\trans{T_k}T_k = \I_{2k}$. As for $\BC$ matrices, the computation speed of a matrix-vector multiplication $Rx$ can be increased significantly by transforming it to the centrosymmetric domain and one can show that~\eqref{eqa:speedup} holds for $n=2$, $p=q$ and $m=t$ (see Appendix~A.2).
\begin{definition}
Let $\mathcal{SCS}(q,t)\subset\R^{2q\times 2t}$ denote the set of \emph{skew-centrosymmetric matrices} of the form
\begin{align}\label{eq:nrs}
D = \left[\begin{array}{c;{4pt/3pt}c}
d_{1} & d_{2}\\[0.3em]
\hdashline[4pt/3pt]\\[-1em]
-J_q d_{2} J_t & -J_q d_{1} J_t
\end{array}\right],\quad\inR{d_i}{p}{t},
\end{align}
with $DJ_{2t}=-J_{2q}D$.
\end{definition}
\noindent In contrast to $\CS$ matrices, a skew-centrosymmetric ($\SCS$) matrix is block anti-diagonalizable by~\eqref{eq:bdiagrefl}, i.e.
\begin{align}\label{eq:bdiagnrefl}
\tilde{D}=\trans{T}_q D T_t =
\begin{bmatrix}
0 & d_1-d_2 J_t\\
d_1+d_2 J_t & 0
\end{bmatrix}.
\end{align}

\section{Main Results}\label{sec:main}
In the following, it will be assumed that the storage ring is divided into $S$ sections of equal length $L/S$. For demonstrating the $\BC$ and the $\CS$ properties of the orbit response matrix, it will be assumed that the $\beta$-function is periodic with period $L/S$ and centrosymmetric with respect to $L/2$, respectively, i.e. $\beta$ can be mirror-reflected about the middle of the storage ring. For our convenience, it will be assumed that $S$, the number of corrector magnets per section and the number of monitors per section are all even.
\subsection{Properties of the Orbit Response Matrix}
\begin{proposition}[Block-Circulant $\mathbf{R}$]\label{thm:bcorm}
Suppose that $\beta(s)=\beta(s+L/S)$ and that each one of the $S$ storage ring sections contains $N_B$ beam position monitors and $N_{C}$ corrector magnets, placed at ring locations $s_1^B,\dots,s_{N_{B}}^B$ and $s_1^C,\dots,s_{N_{C}}^C$, respectively. Suppose that this arrangement is repeated for the $(S-1)$ following sections. Then $\mathbf{R}\in\BC(S,N_B,N_C)$.
\end{proposition}
\begin{IEEEproof}
Partition the first $N_B$ rows into $N_B\times N_C$ blocks. Let  $\sigma(\cdot)$ denote the module operator that is formulated as $\sigma(n+kN_C)=(n+kN_C-1 \mod SN_C)+1$. According to the $\BC$ structure \eqref{eq:bc}, we must show that $\mathbf{R}_{m+kN_B,\sigma(n+kN_C)}=\mathbf{R}_{m,n}$ for $k=1,\dots,S-1$, $m=1,\dots,N_B$ and $n=1,\dots,N_C$. From the definition of $\mathbf{R}_{m,n}$ in~\eqref{eq:ormelem}:
\begin{align*}
&\mathbf{R}_{m+kN_B,\sigma(n+kN_C)}\\
&= \frac{\sqrt{\beta^B_{m+kL/S}\beta^C_{n+kL/S}}}{2\sin(\pi Q)}\cos\left(\pi Q-\abs{\phi^B_{m+kL/S}-\phi^C_{n+kL}}\right)\\
&= \frac{\sqrt{\beta^B_m\beta_n^C}}{2\sin(\pi Q)}\cos\left(\pi Q-\abs{\phi^B_m+k\frac{2\pi Q}{S/L}-\left(\phi_n^C+k\frac{2\pi Q}{S/L}\right)}\right)\\
&= \frac{\sqrt{\beta^B_m\beta_n^C}}{2\sin(\pi Q)}\cos\left(\pi Q-\abs{\phi^B_m-\phi_n^C}\right)=\mathbf{R}_{m,n},
\end{align*}
where we used the fact that $\phi(s+kL)=\phi(s)+k\frac{2\pi Q}{L/S}$ for a periodic $\beta$, which can be verified from~\eqref{eq:phaseadvance}.
\end{IEEEproof}

\begin{proposition}[Centrosymmetric $\mathbf{R}$]\label{thm:reflorm}
Suppose that $\beta(L/2+s)=\beta(L/2-s)$.  In addition, suppose that the position of the monitors and magnets is reflection-symmetric as well, i.e. for each $s_k^X$ there is a $s_p^X$ s.t. $s_p^X=L-s_k^X$ for $X=\lbrace B, C\rbrace$. Then $\mathbf{R}\in\CS\left(SN_B/2, SN_C/2\right)$.
\end{proposition}
\begin{IEEEproof}
The top-half of the matrix must be a vertically and horizontally reflected version of the bottom-half of the matrix. For the top-left and bottom-right sub-blocks of the matrix, $\mathbf{R}_{SN_B/2-n,SN_C/2-m}$ must equal $\mathbf{R}_{SN_B/2+n+1,SN_C/2+m+1}$ for all combinations of $n=0,\dots,\pm SN_B/2-1$ and $m=0,\dots,SN_C/2-1$. After setting $\tilde{s}_k^A=L/2-s^A_{X+k}$ and noting that $\phi(L/2\pm s)=\phi(L/2)\pm\phi(s)$, we obtain:
\begin{align*}
&\mathbf{R}_{SN_B/2-n,SN_C/2-m} =\frac{\sqrt{\beta(L/2-\tilde{s}^B_n)\beta(L/2-\tilde{s}^C_m)}}{2\sin(\pi Q)}\\
&\qquad\times\cos\left(\pi Q-\abs{\phi(L/2-\tilde{s}^B_n)-\phi(L/2-\tilde{s}^C_m)}\right),\\
&=\frac{\sqrt{\beta(L/2\!+\tilde{s}^B_n)\beta(L/2\!+\tilde{s}^C_m)}}{2\sin(\pi Q)}\!\!\underbrace{\cos\left(\pi Q-\abs{\phi(\tilde{s}^C_m)\!-\phi(\tilde{s}^B_n)}\right)}_{=\cos\left(\pi Q-\abs{\phi(L/2+\tilde{s}^C_m)-\phi(L/2+\tilde{s}^B_n)}\right)},\\
&=\mathbf{R}_{SN_B/2+n+1,SN_C/2+m+1},
\end{align*}
and analogously for the top-right and bottom-left sub-blocks.\end{IEEEproof}
Propositions~\ref{thm:bcorm} and~\ref{thm:reflorm} are intuitive results. If the magnetic lattices, beam position monitors and corrector magnets are arranged in a symmetric pattern, then the orbit response matrix inherits the same symmetric pattern. The $\BC$ property means that a circulant shift of $N_B$ and $N_C$ elements can be applied to the beam displacement $\mathbf{y}$ and magnet inputs $\mathbf{u}$ in~\eqref{eq:ormeq} without changing the system behavior, while the $\CS$ property means that each vector can be mirror-reflected about its middle. If Propositions~\ref{thm:bcorm} and~\ref{thm:reflorm} simultaneously hold, the orbit response matrix inherits additional properties, which are summarized in the following proposition. Note that if $\beta$ is periodic with period $L/S$ \emph{and} $\CS$ with respect to $L/2$, then $\beta$ is $\CS$ with respect to the middle of each of the $S$ sections.

\begin{proposition}[Centrosymmetric and Block-Circulant $\mathbf{R}$]\label{thm:reflblocks}
Suppose that the conditions in Propositions~\ref{thm:bcorm} and~\ref{thm:reflorm} all hold, so that $\mathbf{R}\in\BC(S,N_B,N_C)\cap\CS(SN_B/2,SN_C/2)$, and let $\inR{\mathbf{r}_k}{N_B}{N_C}\!$, $k\!=\!0,\dots,S\!\sm\!1$, denote the sub-blocks of $\mathbf{R}$. Then:
\begin{enumerate}
\item $\mathbf{r}_0,\mathbf{r}_{S/2}\in\CS(N_B/2,N_C/2)$
\item $\mathbf{r}_{S/2+k}J_{N_C} = J_{N_B}\mathbf{r}_{S/2-k}$
\end{enumerate}
\end{proposition}
\begin{IEEEproof}
See Appendix~A.3. 
\end{IEEEproof}

\subsection{Decompositions of the Orbit Response Matrix}\label{sec:bcdecom}
The $\BC$ property ensures that $\mathbf{R}$ can be block-diagonalized by pre- and post-multiplication with the discrete Fourier matrix $F_S$ defined in~\eqref{eq:fouriermatrix}. By defining $\mathbf{\hat{y}}\eqdef(F_S^*\otimes\I_{N_B})\mathbf{y}$, $\mathbf{\hat{u}}\eqdef(F_S^*\otimes\I_{N_C})\mathbf{u}$ and $\mathbf{\hat{d}}\eqdef(F_S^*\otimes\I_{N_B})\mathbf{d}$, the dynamics~\eqref{eq:ormeq} can be mapped into the discrete (spatial) Fourier domain as
\begin{align}\label{eq:ormeqfourier}
\mathbf{\hat{y}}(\inv{z}) = \mathbf{\hat{R}}g(\inv{z}) \mathbf{\hat{u}}(\inv{z}) +\mathbf{\hat{d}}(\inv{z}),
\end{align}
where $\mathbf{\hat{R}}\eqdef\diag\{\mathbf{\hat{R}}_0,\dots,\mathbf{\hat{R}}_{S-1}\}$. When a vector is mapped into the Fourier domain as in $\mathbf{\hat{y}}=(F_S^*\otimes\I_{N_B})\mathbf{y}$, the Kronecker product between $F_S^*$ and $\I_{N_B}$ means that the $k$th displacements of each superperiod are grouped. The Fourier transform is then applied to equidistant samples at $s^B_m + k L$, $k=0,\dots,S-1$. This yields the Fourier coefficients for the spatial frequencies $\omega_k = 2\pi k/S$. The block-diagonal structure of $\mathbf{\hat{R}}$ means that the spatial Fourier coefficients of the displacements at frequency $\omega_k$ are not modified by magnetic inputs at frequency $\omega_j$ for $n\neq j$. The Fourier coefficients are, however, influenced by other Fourier coefficients of the same spatial frequency that have a different starting point $s_n^C$ for the equidistant samples $s_n^C+kL$. Note that one could apply the spatial Fourier transform to any matrix, but the resulting $\mathbf{\hat{R}}$ is block-diagonal if and only if $\mathbf{R}$ is $\BC$~\cite{CIRCBOOK}.

Analogously to the $\BC$ case, the dynamics~\eqref{eq:ormeq} can be mapped to the $\CS$ domain by defining $\mathbf{\tilde{y}}\eqdef\trans{T}_{SN_B/2}\mathbf{y}$, $\mathbf{\tilde{u}}\eqdef\trans{T}_{SN_C/2}\mathbf{u}$ and $\mathbf{\tilde{d}}\eqdef\trans{T}_{SN_C/2}\mathbf{d}$. The resulting $\mathbf{\tilde{R}}\eqdef\trans{T}_{SN_B/2} \mathbf{R} T_{SN_C/2}$ is block-diagonal if and only if $\mathbf{R}$ is $\CS$. The transformation $\mathbf{\tilde{y}}=\trans{T}_{SN_B/2}\mathbf{y}$ groups elements $k$ and $k+SN_C/2$ of $\mathbf{y}$ and assigns their sum and differences to $\mathbf{\tilde{y}}$. The block-diagonalized $\mathbf{\tilde{R}}$ reflects the fact that the sum (difference) of the displacements, is solely modified by the sum (difference) of the magnetic kicks.

When $\mathbf{R}$ is \emph{both} $\BC$ and $\CS$, the matrices $\mathbf{\hat{R}}$ as well as $\mathbf{\tilde{R}}$ can be further decomposed. For the decomposition of $\mathbf{\hat{R}}$, we start by rewriting the complex-valued blocks of  the $\BC$ decomposition $\mathbf{\hat{R}}$ using~\eqref{eq:diagblocksB} as
\begin{align*}
\mathbf{\hat{r}}_n &=\mathbf{r}_0 + (-1)^n \mathbf{r}_{S/2}\\&\qquad + \sum_{k=1}^{S/2-1} \left( \mathbf{r}_k e^{-i\frac{2\pi nk}{S}} + J_{N_B}\mathbf{r}_{k} J_{N_C} e^{i\frac{2\pi nk}{S}} \right),
\end{align*}
where the second part of Proposition~\ref{thm:reflblocks} was used after reformulating it as $\mathbf{r}_{S-k}=J_{N_B}\mathbf{r}_{k} J_{N_C}$. Separating the real and imaginary parts of $\mathbf{\hat{r}}_n$ yields
\begin{align*}
\Real{\mathbf{\hat{r}}_n} &= \mathbf{r}_0 + (-1)^n \mathbf{r}_{S/2}\\&\phantom{=} + \sum_{k=1}^{S/2-1} \cos\left(\frac{2\pi nk}{S}\right)\left( \mathbf{r}_k + J_{N_B}\mathbf{r}_{k} J_{N_C} \right),\\
\Imag{\mathbf{\hat{r}}_n}\!&=\!\!\sum_{k=1}^{S/2-1}\!\! \sin\left(\frac{2\pi nk}{S}\right)\left( J_{N_B}\mathbf{r}_{k} J_{N_C}-\mathbf{r}_k \right).
\end{align*}
Common to matrices with symmetric structures, such as $\BC$ or $\CS$ matrices, is that they form an algebra (see Appendix~\ref{app:matrices}). Because $\mathbf{r}_0$, $\mathbf{r}_{S/2}$ and $\mathbf{r}_k + J_{N_B}\mathbf{r}_{k} J_{N_C}${\footnote{\label{note1}This can be shown by pre- and post-multiplication with $J_{N_B}$ and $J_{N_C}$, respectively.}} are $\CS$, the real part of $\mathbf{\hat{R}}_n$ is $\CS$, while the imaginary part is $\SCS$, because $J_{N_B}\mathbf{r}_{k} J_{N_C}-\mathbf{r}_k$ is $\SCS$\footnotemark[1]. Each of the Fourier blocks  $\mathbf{\hat{R}}_n$ can therefore be pre- and post-multiplied by $\trans{T}_{N_B/2}$ and $T_{N_C/2}$, which will separate the real and imaginary part, because $\CS$ matrices are block-diagonalized while $\SCS$ matrices are block anti-diagonalized by the transformation~\eqref{eq:centromatrix}.

The decomposition of the $\CS$ decomposition  $   \mathbf{\tilde{R}}$ can be found in Appendix~A.4, in which it is shown that, if $\mathbf{R}$ is $\BC$ as well, each of the blocks of $\mathbf{\tilde{R}}$ is $\CS$ and can therefore be decomposed using~\eqref{eq:bdiagrefl}. Note that the $\BC$ structure is a more stringent requirement than needed, i.e. the doubly $\CS$ decomposition only requires that the $\beta$-function is $\CS$ with respect to $L/4$. Table~\ref{tab:allresults} summarizes the results from this section and characterizes the $\BC$, $\CS$ and their further decompositions by showing the formulae for block-diagonalization and the resulting shapes of the block-diagonalized matrices. The table also addresses symmetric approximations of $\mathbf{R}$, which are treated in the following section.
\ifTwoColumn
	\begin{table*}[htpb]
\else
	\begin{sidewaystable}[htpb]
\fi
\caption{Symmetric Decompositions for $S=6$}
\begin{tabular}
	{|p{0.11\textwidth}
	|>{\centering\arraybackslash}p{0.41\textwidth}
	|>{\centering\arraybackslash}p{0.41\textwidth}|}
\hline\TBstrut
Decomposition & Centrosymmetric ($\CS$) & Decomposition of the $\CS$ decomposition ($\CS-\BC$)\Bstrut\\
\hline
Diagonalization ($\mathbf{\hat{R}}$)
& $\trans{T}_{SN_B/2}\mathbf{R}T_{SN_C/2}$
& $\left(\I_2\otimes \trans{T}_{SN_B/4}\right)\trans{T}_{SN_B/2}\mathbf{R}T_{SN_C/2}\left(\I_2\otimes T_{SN_C/4}\right)$\TBLstrut\\
\hline
Approximation$^{\dagger}$  ($\mathbf{R}^\star$)
& $\frac{1}{2}\left(\mathbf{R}^p + J_{SN_B} \mathbf{R}^p J_{SN_C}\right)$
& $\frac{1}{2S}\sum_{k=0}^{S-1}
\left(\Omega_S^k\!\otimes\!\I_{N_B}\right)^\Tr\!
\left(\mathbf{R}^p\!+\!J_{SN_B} \mathbf{R}^p J_{SN_C}\right)
\left(\Omega_S^k\!\otimes\!\I_{N_C}\right)$\TBLstrut\\
\hline
\raisebox{-30pt}{Shape of $\mathbf{\hat{R}}$} &
\raisebox{-0.95\totalheight}{\includegraphics[scale=1]{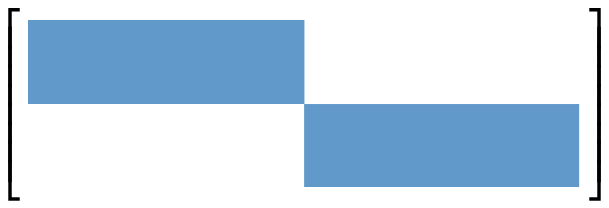}}
&
\raisebox{-0.95\totalheight}{\includegraphics[scale=1]{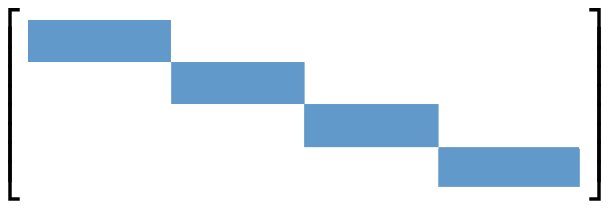}}\\
\hline
\raisebox{-30pt}{Shape of $\hat{\Delta}$} &
\raisebox{-0.95\totalheight}{\includegraphics[scale=1]{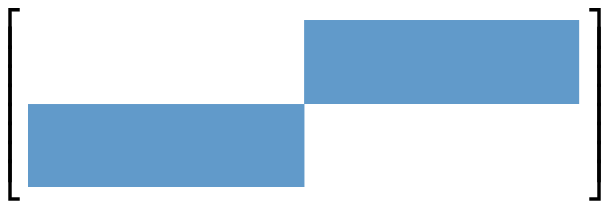}}&
\raisebox{-0.95\totalheight}{\includegraphics[scale=1]{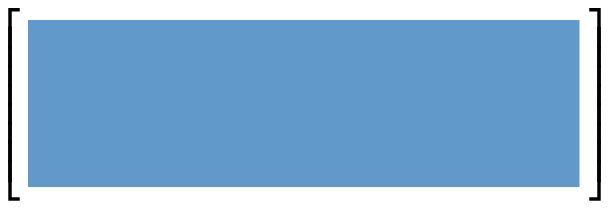}}\Bstrut\\
\hline
\multicolumn{3}{p{400pt}}{\phantom{space}}\\
\hline\TBstrut
Decomposition & Block-Circulant ($\BC$) & Decomposition of the $\BC$ decomposition ($\BC-\CS$)\Bstrut\\
\hline
Diagonalization ($\mathbf{\hat{R}}$)
& $\left(F_S^*\otimes \I_{N_B}\right) \mathbf{R} \left(F_S\otimes\I_{N_C}\right)$
& $\left(\I_S\otimes \trans{T}_{N_B/2}\right)\left(F_S^*\otimes\I_{N_B}\right) \mathbf{R} \left(F_S\otimes \I_{N_C}\right)\left(\I_S\otimes T_{N_C/2}\right)$ \TBLstrut\\
\hline
Approximation$^{\dagger}$ ($\mathbf{R}^\star$)
&$\!\!\frac{1}{S}
\sum_{k=0}^{S-1}\left(\Omega_S^k\otimes\I_{N_B}\right)^\Tr
\mathbf{R}^p\left(\Omega_S^k\otimes\I_{N_C}\right)$
& $\frac{1}{2S}\sum_{k=0}^{S-1}\left(\Omega_S^k\!\otimes\!\I_{N_B}\right)^\Tr\!\left(\mathbf{R}^p\!+\!J_{SN_B} \mathbf{R}^p J_{SN_C} \right)
\left(\Omega_S^k\!\otimes\!\I_{N_C}\right)$\TBLstrut\\
\hline
\raisebox{-30pt}{Shape of $\mathbf{\hat{R}}$} &
\raisebox{-0.95\totalheight}{\includegraphics[scale=1]{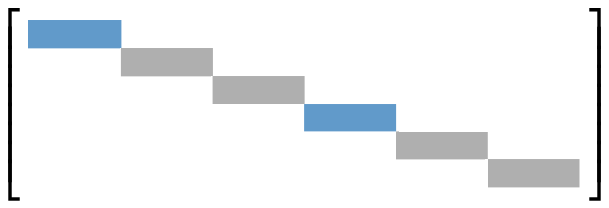}}&
\raisebox{-0.95\totalheight}{\includegraphics[scale=1]{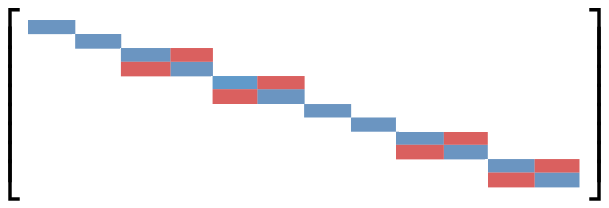}}\\
\hline
\raisebox{-30pt}{Shape of $\hat{\Delta}$} &
\raisebox{-0.95\totalheight}{\includegraphics[scale=1]{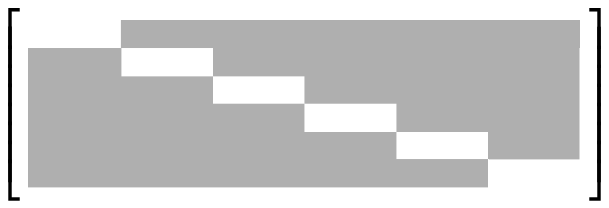}}&
\raisebox{-0.95\totalheight}{\includegraphics[scale=1]{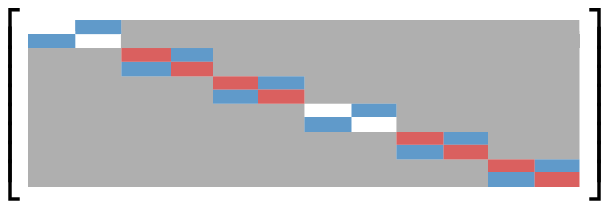}}\\
\hline
\end{tabular}
\hspace{3ex}~Blue, red and gray blocks refer to real, purely imaginary and complex-valued numbers, respectively. ~ $^{\dagger}$The matrix $\Omega_S$ is defined in~\eqref{eqa:circperm}.
\label{tab:allresults}
\ifTwoColumn
	\end{table*}
\else
	\end{sidewaystable}
\fi

\section{Broken Symmetry}\label{sec:brokensymmetry}
In practice, the regular arrangement of magnetic lattices, beam position monitors and corrector magnets is compromised by space constraints, e.g. there will be one section where the injection device -- the entry point for the electrons -- will need to be fitted. This will lead to an asymmetric placement of one or more of the aforementioned components. It can also be that some of the magnets in the magnetic lattices are not perfectly aligned to the vertical or horizontal plane. This, in turn, leads to an asymmetry of the $\beta$-function. Because the symmetry of the orbit response matrix is solely based on the symmetry of the betatron function as well as on the placement of monitors and magnets, the orbit response matrix will inherit any asymmetry. If we are to exploit the symmetric structure of the synchrotron for the controller, then the matrix $\mathbf{K}$ in~\eqref{eq:ctrmat} must have the symmetry properties. If $\mathbf{R}$ is $\BC$ and/or $\CS$, the gain matrix $\mathbf{K}$, which can also be computed from $\mathbf{K}=\inv{(\trans{\mathbf{R}}\mathbf{R}+\mu\I_{SN_B})}\trans{\mathbf{R}}$, will necessarily have the symmetry properties because each of our symmetric structures form an algebra.

\subsection{Symmetric Approximations}\label{sec:approx}
Given a perturbed orbit response matrix $\mathbf{R}^p$ that does \emph{not} satisfy the symmetry conditions, a matrix $\mathbf{R}^\star$ is sought that approximates $\mathbf{R}^p$ and features the $\BC$ and/or $\CS$ properties. This problem can be formulated as an optimization problem,
\begin{align}\label{eq:ormapproxsec}
\mathbf{R}^\star = \argmin_{X\in\,\mathcal{S}} \frobnorm{X-\mathbf{R}^p}^2,
\end{align}
where $\mathcal{S}\in\left\lbrace	\BC,\CS,\BC\cap\CS\right\rbrace$ and the Frobenius norm was used because this norm leads to closed-form solutions. If temporal dynamics were involved, a more appropriate choice would be the $\infty$-norm, for which $\mathbf{R}^\star$ would also reflect the stability properties of $\mathbf{R}^p$\footnote{The resulting problem would be a linear program.}. In~\cite{CIRCAPPROX}, a solution is derived for the case that $\mathcal{R}_p$ is a Toeplitz matrix and in Appendix~\ref{app:approx}, the proofs are extended for a general~$\mathcal{R}_p$ and $\mathcal{S}\in\left\lbrace\BC,\CS,\BC\cap\CS\right\rbrace$. The results obtained are summarized in Table~\ref{tab:allresults}. They essentially consist of averaging over the sub-blocks of $\mathbf{R}^p$ according to the corresponding structure of $\mathcal{S}$, e.g. when $\mathcal{S}=\BC$ the diagonal block $r_0^\star$ is obtained from averaging over all sub-blocks of $\mathbf{R}^p$ that are lying on the diagonal.

\subsection{Approximation Error}\label{sec:uncertainty}
When the gain matrix $\mathbf{K}$ is computed using an approximation $\mathbf{R}^\star$, the stability properties of the resulting closed-loop system might be affected, i.e. the system might be stable if $\mathbf{K}$ is computed using $\mathbf{R}^p$ but unstable when computed using $\mathbf{R}^\star$. To measure the amount of asymmetry, an approximation error is defined as $\Delta \eqdef \mathbf{R}^p-\mathbf{R}^\star$. For a symmetric approximation problem, such as~\eqref{eq:ormapproxsec}, the structure of $\Delta$ can be determined from transforming the optimization~\eqref{eq:ormapproxsec} into the symmetric domain $\mathcal{S}$, i.e. by rewriting the norm in~\eqref{eq:ormapproxsec} using the symmetric transformations $\mathcal{T}_{\mathcal{S},l},\mathcal{T}_{\mathcal{S},r}$ as
\begin{align}\label{eq:normsym}
\frobnorm{\herm{\mathcal{T}}_{\mathcal{S},l}\left(X-\mathbf{R}^p\right)\mathcal{T}_{\mathcal{S},r}}=\frobnorm{\tilde{X}-\mathbf{\tilde{R}}^p_\pl-\mathbf{\tilde{R}}^p_\bot},
\end{align}
where $\tilde{X}=\herm{\mathcal{T}}_{\mathcal{S},l}X\mathcal{T}_{\mathcal{S},r}$, $\mathbf{\tilde{R}}^p_\pl+\mathbf{\tilde{R}}^p_\bot = \herm{\mathcal{T}}_{\mathcal{S},r}\mathbf{R}^p\mathcal{T}_{\mathcal{S},r}$ and where $\mathbf{\tilde{R}}^p_\pl$ has the same structure as $\tilde{X}$ such that
\begin{align}\label{eq:complement}
\rre\left(\mathbf{\tilde{R}}^p_\bot\right)\circ\rre\left(\mathbf{\tilde{R}}^p_\pl\right) = 0,\quad
\iim\left(\mathbf{\tilde{R}}^p_\bot\right)\circ\iim\left(\mathbf{\tilde{R}}^p_\pl\right) = 0.
\end{align}
Note that the norm is invariant w.r.t. multiplication with an orthonormal matrix~\cite[Ch. 2.3.5, p.~75]{GOLUB4}. From~\eqref{eq:normsym}, it becomes clear that $\mathbf{\tilde{R}}^\star=\mathbf{\tilde{R}}^p_\pl$ and $\tilde{\Delta}=\mathbf{\tilde{R}}^p_\bot$ and the solution to~\eqref{eq:ormapproxsec} could be found by setting $\mathbf{R}^\star=\mathcal{T}_{\mathcal{S},l}\mathbf{\tilde{R}}^\star\herm{\mathcal{T}}_{\mathcal{S},r}$. The shapes of $\tilde{\Delta}$ for $\mathcal{S}\in\lbrace \BC,\CS,\BC\cap\CS\rbrace$ are depicted in Table~\ref{tab:allresults}. Note that for the doubly $\CS$ decomposition (column $\CS-\BC$ in Table~\ref{tab:allresults}) we use the approximation that yields $\mathbf{R}^\star\in\BC\cap\CS$ and~\eqref{eq:complement} therefore does not hold.

\section{Case Study: Diamond-II}\label{sec:casestudy}
Diamond Light Source is the UK’s national synchrotron facility, which produces synchrotron light for research.  The Diamond-I synchrotron is a \nth{3}-generation light source in which electrons circulate around the $560$~m storage ring at an energy of $3$~GeV. The upcoming Diamond-II conversion will upgrade the synchrotron to a \nth{4}-generation light source that operates at $3.5$~GeV and introduce various changes, such as new lattice technologies and a new orbit feedback system~\cite{DIAMONDII}. At Diamond-I, the orbit feedback system uses $172$ monitors to operate $2\times 173$ corrector magnets -- one set for the vertical and one for the horizontal plane -- at a frequency of $10$~kHz. The feedback reduces the trajectory error of the electrons to 5~$\mu$m in the horizontal and 600~nm in the vertical plane~\cite{PHDSANDIRA}. Diamond-II will use $252$ monitors and $2\times 396$ corrector magnets that will be operated at a frequency of $100$~kHz.
\begin{figure}[htpb]
  	\centering
	\centerline{\includegraphics[scale=1]{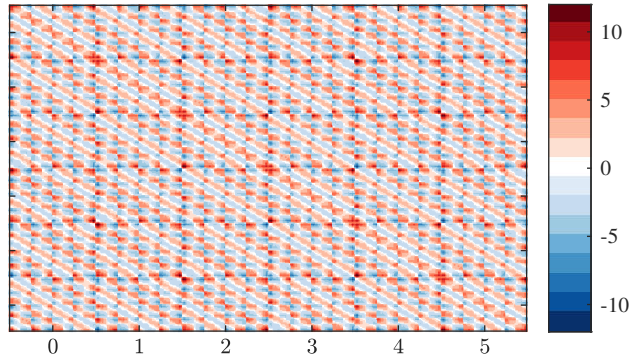}}
  	\caption{Diamond-II orbit response matrix with $S=6$ sections.}
  	\label{fig:orm}
\end{figure}
\subsection{Symmetric Approximations}\label{sec:Diamondapprox} The Diamond-II storage ring will be arranged in $S=6$ superperiods and for this case study we are using a preliminary version of the orbit response matrix, which is shown in Fig.~\ref{fig:orm} for the vertical plane. The $\BC$, $\CS$ and $\BC\cap\CS$ approximations were applied to the orbit response matrix of Fig.~\ref{fig:orm} and the 2-norm ($\twonorm{X}$), the average absolute ($\avg_{ij}(\abs{X_{ij}})$) and the maximum absolute magnitude ($\max_{ij}(\abs{X_{ij}})$), respectively, of the resulting approximation errors are compared to $\mathbf{R}^p$ in Table~\ref{tab:approxerror}. The small errors for the $\BC$ approximation show that the $\BC$ symmetry property is an accurate assumption for $\mathbf{R}^p$, whereas the $\CS$ and $\BC\cap\CS$ approximations yield significantly larger errors. Their maximum singular values, however, are substantially smaller than the maximum singular value of $\mathbf{R}^p$.
\begin{table}[htpb]
\centering
\caption{Approximation Error}
\label{tab:approxerror}
\begin{tabular}{|p{50pt}|p{47pt}|p{50pt}|p{53pt}|}
\hline\TBstrut
$X$&
$\twonorm{X}$&
$\avg_{ij}(\abs{X_{ij}})$&
$\max_{ij}(\abs{X_{ij}})$\Bstrut\\
\hline
$\mathbf{R}^p$&
644&
2.7112&
11.8176\Tstrut\\
\hline
$\Delta$-$\BC$&
0.0355&
0.0001&
0.0009\Tstrut\\
\hline
$\Delta$-$\CS$&
50&
0.1378&
1.4834\Tstrut\\
\hline
$\Delta$-$\BC\cap\CS$&
50&
0.1378&
1.4835\Tstrut\\
\hline
\end{tabular}
\label{tab:approxerror}
\end{table}

\subsection{Orbit Feedback Controller}\label{sec:controller}
\begin{figure}
\centering
\begin{tikzpicture}[align=center,node distance=0.5cm,auto,
	    every node/.style={inner sep=2pt,rectangle, minimum height=1em, text centered},
	    block/.style={draw,inner sep=2pt},
	    sum/.style={draw,minimum height =8pt,circle,inner sep=0pt},
	    nnode/.style={draw,fill=black,minimum height=2pt,circle,inner sep=0pt}]

\node[sum] (min) [] {\sst{$\,\mathbf{-}$}};
\node[block] (K) [right =0.75cm of min] {{$\mathbf{K}c(z^{\sm 1})$}};
\node[block] (P) [right =1.5cm of K] {{$\mathbf{R}^p g(z^{\sm 1})$}};
\node[sum] (plus) [right =0.75cm of P] {\sst{$\mathbf{+}$}};
\node[nnode] (dot) [right =0.25cm of plus] {};
\node[] (y) [right =0.5cm of dot] {{$\mathbf{y}(z^{\sm 1})$}};
\node[] (d) [above =0.75cm of plus] {{$\mathbf{d}(z^{\sm 1})$}};
\node[] (corner) [below =1.25cm of dot] {};

\draw[-Latex, line width=\lw] (min.east) -- (K.west);
\draw[-Latex, line width=\lw] (K.east) -- (P.west);
\draw[-Latex, line width=\lw] (P.east) -- (plus.west);
\draw[-Latex, line width=\lw] (plus.east) -- (y.west);
\draw[-Latex, line width=\lw] (d.south) -- (plus.north);
\draw[line width=\lw] (dot.center) -- ([yshift=-0.5\lw]corner.center);
\draw[-Latex, line width=\lw] (corner.center) -| (min.south);
\node[] () [above=0.05cm of $(K)!0.5!(P)$] {{$\mathbf{u}(z^{\sm 1})$}};
\end{tikzpicture}
\caption{Feedback system.}\label{fig:fbsystem}
\end{figure}
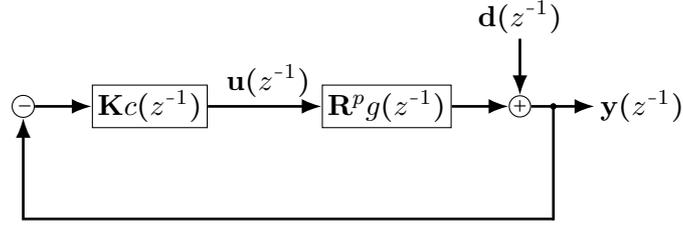

As a proof of concept for Diamond-II, we are interested in how the standard controller $\mathbf{u}(\inv{z})=\mathbf{K}c(\inv{z})\mathbf{y}(\inv{z})$ performs when $\mathbf{K}$ is obtained using a symmetric approximation, i.e. $\mathbf{K}=\inv{(\trans{{\mathbf{R}^\star}}{\mathbf{R}^\star}+\mu\I_{SN_B})}\trans{{\mathbf{R}^\star}}$, while the process model is given by the asymmetric Diamond-II orbit response matrix $\mathbf{R}^p$. We will focus on the orbit correction for the vertical plane, but the results are comparable for the horizontal plane. Fig.~\ref{fig:fbsystem} shows the control system in its standard configuration. It is assumed that all actuators have the same dynamics $g(\inv{z})$,
\begin{align*}
g(\inv{z})=z^{-d}\frac{b_0+b_1 z^{-1}}{1-az^{-1}},
\end{align*}
where $d=7$ is the delay in terms of time steps and the parameters $b_0,b_1$ and $a$ can be found in~\cite[p.~207]{SANDIRACONTROLDESIGN}. An internal model controller is used to form the dynamic part of the controller $c(\inv{z})$~\cite[eq.~(20)]{SANDIRACONTROLDESIGN}.

\begin{figure}[t]
  	\centering
	\includegraphics[scale=1]{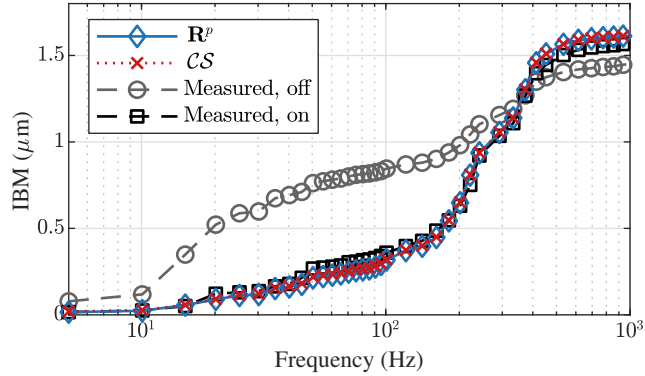}
  	\caption{Simulation of the controller. Also shows the augmented Diamond-I measurements for enabled and disabled feedback.}
  	\label{fig:sim_results_wide}
\end{figure}

The simulation requires the disturbances $\mathbf{d}(t)$ as an input. Because no such measurements are available for Diamond-II yet, the measurements from Diamond-I are used. The disturbance vector is augmented to fit the dimensions of Diamond-II. First, the $80=252-172$ monitor outputs are copied and appended to the $172$ measurements. Second, the augmented disturbances are transformed into mode-space using an SVD of the Diamond-II orbit response matrix and the power spectrum of the modes plotted, such as in~\cite[Ch. 3.5, pp. 68--72]{PHDSANDIRA}. Third, the disturbance spectrum is scaled to obtain a power spectrum comparable to~\cite[Fig.~3.11]{PHDSANDIRA}, where the modes associated to large-magnitude singular values show a larger amplitude. The resulting disturbance profile is depicted in Fig.~\ref{fig:sim_results_wide} (labeled by Measured, off).

\begin{figure}[t]
  	\centering
	\includegraphics[scale=1]{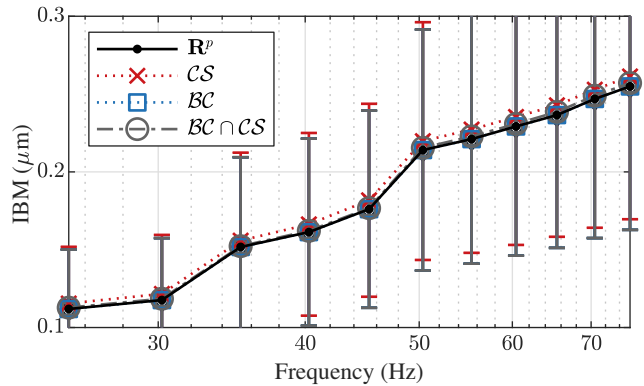}
  	\caption{Close-up of Fig.~\ref{fig:sim_results_wide} with the $\BC$ and $\BC\cap\CS$ approximations.}
  	\label{fig:sim_results_zoom}
\end{figure}

The performance of the controller is measured using the
integrated beam motion (IBM), which is defined as the square root of $\sum_{f=0}^F \frac{2}{F^2}\abs{y(f)}^2$, where $y(f)$ is the discrete Fourier transform of a monitor output and $F$ the frequency in Hz. Fig.~\ref{fig:sim_results_wide} shows the average IBM across all beam position monitors of the storage ring for $\mathbf{K}$ computed using $\mathbf{R}^p$ and the $\CS$ approximation of $\mathbf{R}^p$. For clarity, the simulation results for the $\BC$ and $\BC\cap\CS$ approximations are omitted in Fig.~\ref{fig:sim_results_wide}, but are shown in the close-up in Fig.~\ref{fig:sim_results_zoom}. The results show that the controller performs only slightly worse when a symmetric approximation is used. In Fig.~\ref{fig:sim_results_zoom}, it can be seen that the $\CS$ approximation yields a slightly larger average trajectory error, which is related to the large approximation error.

The nominal stability of the controller, i.e. when no uncertainty is present, can be verified by calculating the poles of the closed-loop transfer functions. The closed-loop transfer functions of the system in Fig.~\ref{fig:fbsystem} are given by
\begin{subequations}\begin{align}
\mathbf{y}(\inv{z}) &= \inv{\left(\I_{N_B}+\mathbf{R}^p\mathbf{K}L(\inv{z}) \right)}\mathbf{d}(\inv{z}),\label{eq:Sy}\\
\mathbf{u}(\inv{z}) &= \mathbf{K}c(\inv{z})\inv{\left(\I_{N_C}+\mathbf{R}^p\mathbf{K}L(\inv{z}) \right)}\mathbf{d}(\inv{z}),\label{eq:Su}
\end{align}\end{subequations}
where $L(\inv{z}) = g(\inv{z})c(\inv{z})$. When the controller uses a symmetric approximation, the control system cannot be diagonalized by an SVD of $\mathbf{R}^p$, such as in~\cite{SANDIRACONTROLDESIGN}. One can, however, use an eigendecomposition of $\mathbf{R}^p\mathbf{K}$ to diagonalize~\eqref{eq:Sy} and~\eqref{eq:Su}. E.g. for~\eqref{eq:Su}, one can substitute $\mathbf{R}^p\mathbf{K}=V D \inv{V}$, where $D$ is the diagonal matrix of eigenvalues and the columns of $V$ are the corresponding eigenvectors, which yields
\begin{align*}
\mathbf{u}=\mathbf{K}V\inv{\left(\I_{N_C}+DL(\inv{z}) \right)}c(\inv{z})\inv{V}\mathbf{d}(\inv{z}),
\end{align*}
and the stability can be verified by computing the poles of the diagonal matrix
\ifTwoColumn $
\else \[ \fi
\inv{\left(\I_{N_C}+DL(\inv{z}) \right)}c(\inv{z}).
\ifTwoColumn $
\else \] \fi
The feedback system is stable if all poles of~\eqref{eq:Sy} and~\eqref{eq:Su} have magnitude smaller than 1, which is satisfied for $\mathbf{K}$ computed using $\mathbf{R}^p$ as well as using the symmetric approximations.

\subsection{Benchmarks on Hardware}
The Diamond-I feedback system is implemented on $24$ processors, which are distributed around the storage ring and require a complex distributed network topology, while for Diamond-II the computations will be centralized and the feedback system will be implemented on one multicore processor~\cite{TIC6678}. The new setup simplifies the network topology but increases the performance requirements on the multicore processor. The clock-frequency of the processor is $1.4$~GHz and the targeted operating frequency of $100$~kHz therefore allows for $14,000$ processor cycles. Without decomposition, the controller computations require $252\times 396\approx 100,000$ multiply-accumulate operations. For demonstrating the speed advantages of the decomposition, the matrix-vector multiplication required by the controller has been implemented on the processor. When a decomposition is used, the control input is computed as
\begin{align}\label{eq:symgain}
\mathbf{u}(\inv{z})=\mathcal{T}_{\mathcal{D},l}\,\mathbf{\tilde{K}}\,\trans{\mathcal{T}}_{\mathcal{D},r}\,c(\inv{z})\mathbf{y}(\inv{z}),
\end{align}
where $\mathcal{D}\in\lbrace \CS, \BC, \CS\!\sm\!\BC,\BC\!\sm\!\CS\rbrace$ refers to the decompositions in Table~\ref{tab:allresults} and $\mathcal{T}_{\mathcal{S},l},\mathcal{T}_{\mathcal{S},r}$ to the corresponding transformation, and $\mathbf{\tilde{K}}$ is the decomposed gain matrix.

\begin{figure}[t]
  	\centering
	\includegraphics[scale=1]{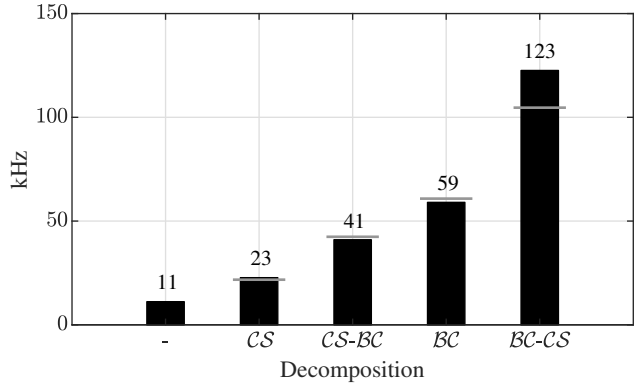}
  	\caption{Performance measurement of the controller. The leftmost timing does not use any decomposition. The gray bars refer to the theoretical speed-up.}
  	\label{fig:speed}
\end{figure}

Fig.~\ref{fig:speed} shows the results that were obtained for the implementation of~\eqref{eq:symgain} on a single core of the processor. The performance is measured as $1/t$, where $t$ is the time required to execute one matrix-vector multiplication and the horizontal gray bars refer to the theoretical speed-up that was calculated using~\eqref{eqa:speedup} and the computation frequency of $11$~kHz, which corresponds to the frequency of the matrix-vector multiplication without decomposition. Using the $\BC\!\sm\!\CS$ decomposition, the computation frequency is more than eleven times faster than without decomposition and beyond the targeted operating frequency of $100$~kHz. The results also show that for the $\BC\!\sm\!\CS$ decomposition the speed-up of the implementation is significantly larger than the theoretical prediction. The reason is that, in addition to the number of operations required, the performance of the processor is also limited by memory operations, e.g. the time needed to transport the matrix data from the memory to the core. The reduced memory requirements of the decomposition therefore indirectly benefit the computation time.
\ifTwoColumn
\newpage
\fi
\section{Conclusions}
In this paper, we have shown that the orbit response matrix of a synchrotron inherits the mirror-reflective and periodic properties of the betratron function. The mirror-reflective and periodic properties of the betatron function manifest themselves in a centrosymmetric and block-circulant structure of the orbit response matrix, which results in a controller gain matrix that has the same structure. These structural symmetries can be used to decompose the gain matrix and perform the matrix-vector multiplication, which is required for computing the control inputs, in the symmetric domain. For the symmetries discussed in this paper, the transformation of a vector into the symmetric domain is computation-efficient and the matrix-vector multiplication in the symmetric domain requires far less multiply-accumulate operations than in the original domain.

In practice, the mirror-reflective and periodic properties of the betatron function are only approximate, e.g. a symmetry might be broken at some point around the storage ring due odd placements of monitors or magnets. The asymmetry of the betatron function is inherited by the orbit response matrix, which, in turn, results in an asymmetric controller gain matrix that cannot be block-diagonalized using a symmetric decomposition. To recover the symmetric structure, an optimization problem was formulated in which a matrix was sought that has the corresponding symmetry properties and approximates the asymmetric orbit response matrix. We presented closed-form solutions for the centrosymmetric, block-circulant and combined symmetry cases and an alternative approach that finds the solution by transforming the problem into the symmetric domain. This alternative approach showed that the approximation error-matrix always has a particular structure, which complements the approximation in the symmetric domain.

We concluded the paper with a case study of the Diamond-II synchrotron. We approximated the Diamond-II orbit response matrix for the centrosymmetric, block-circulant and combined symmetry cases and the approximation errors showed that the block-circulant approximation yields a small error, while the centrosymmetric and combined symmetry cases yielded larger errors. The controller gain matrix was computed using the different approximations and the closed-loop simulations showed that there is only a minuscule difference in trajectory error correction between the gain matrices computed using the approximations and using the asymmetric orbit response matrix. The nominal stability for all approximations was verified using an eigendecomposition of the closed-loop transfer functions. The robust stability analysis using the structured approximation error is currently being considered.

We completed the case study using a single-core implementation of the controller computations and demonstrated the significant improvement of the computation-speed when the matrix-vector multiplication is carried out in the symmetric domain. For the combined symmetry case, the controller computations in the symmetric domain were more than ten times faster than in the original domain  and already beyond the targeted operating frequency of $100$~kHz. It is expected that, after parallelizing the matrix-vector multiplication, the computations will be fast enough to enable the use of more advanced control algorithms, such as algorithms that require real-time optimization and explicitly consider system constraints, such as amplitude and slew-rate constraints of the magnets.
\bibliographystyle{IEEEtran}
\bibliography{master_bib}
\appendices
\renewcommand\thesubsectiondis{\thesection.\arabic{subsection}}
\section{}\label{app:matrices}
\subsection{Block-Circulant Matrices}\label{sec:bcmatrices}
For $\BC$ matrices it holds that~\cite{CIRCBOOK}
\begin{align}\label{eqa:circperm}
B\left(\Omega_n\otimes\I_m\right)=\left(\Omega_n\otimes\I_p\right) B,\,\,\Omega_n = {\arraycolsep=3pt%
\left[
  \begin{array}{@{}*2c}
0 & \I_{n-1}\\ 1 & 0
  \end{array}
\right]},
\end{align}
where $\Omega_n$ is the cyclic shift matrix with $\trans{\Omega}_n\Omega_n=\I_n$ and $\Omega^n_n=\I_n$. Using~\eqref{eqa:circperm}, it can be shown that $\BC$ matrices form an algebra~\cite{CIRCBOOK}, i.e. that sums and products of $\BC$ matrices yield another $\BC$ matrix. Any $\BC$ matrix can be represented as~\cite{CIRCBOOK}
\begin{align}\label{eqa:bc2}
B = \sum_{k=0}^{n-1}\Omega_n^k\otimes b_k.
\end{align}
When the $\BC$ matrix $B$ in~\eqref{eq:bdiagB} is real, the blocks $\nu_j$ possess additional structure that is inherited from the Fourier matrix $F_n$. If $n$ is even, $\nu_0$ and $\nu_{n/2}$ are real while for $i=1,\dots,n/2\!\sm\!1$ it holds that $\nu_i=\bar{\nu}_{n-i}$. If $n$ is odd, the only real-valued block is $\nu_0$ and the latter holds for $i=1,\dots,(n\sm1)/2$. The same pattern of complex conjugates is exploited during a Fast Fourier Transformation and, according to the properties of $\nu_j$, only the first $n/2$ blocks must be considered for a matrix-vector multiplication.

\subsection{Centrosymmetric and Skew-Centrosymmetric Matrices}\label{sec:reflmatrices}
As for $\BC$ matrices, it can be shown that $\CS$ and $\SCS$ matrices also form algebras~\cite{CENTROSYM}. By reversing the order of the second half of the rows and columns, a $\CS$ matrix can be permuted into a $\BC$ matrix of order 2:
\begin{lemma}\label{thm:perm}
The permutation matrices $P_l=\diag(\I_{q},J_{q})$ and $P_r=\diag(\I_{t},J_{t})$ permute $R\in\CS(q,t)$ into a $\BC$ matrix of order $n=2$:
\begin{align*}
P_l R P_r = \left[\begin{array}{c;{4pt/3pt}c}
r_{1} & r_{2}J_t\\[0.3em]
\hdashline[4pt/3pt]\\[-1em]
r_{2}J_t & r_{1}
\end{array}\right]\in\BC(2,q,t).
\end{align*}
\end{lemma}
\begin{IEEEproof}
Evaluating the product $P_l R P_r$ yields the result.
\end{IEEEproof}
Lemma~\eqref{thm:perm} shows that $\CS$ matrices and $\BC$ matrices are closely related. For our purpose, it is sufficient to use Lemma~\eqref{thm:perm} to show that~\eqref{eqa:speedup} holds for $n=2$.

\subsection{Proof of Proposition~\ref{thm:reflblocks}\label{app:bccs}}
To find the algebraic conditions a simultaneously $\BC$ and $\CS$ matrix satisfies, consider the permutation matrix $J_n$ acting on the cyclic shift matrix $\Omega_n$
\begin{align}\label{eq:jomega}
J_n\Omega_n J_n = \begin{bmatrix}
0 & 1\\ \I_{n-1} & 0\end{bmatrix} = \Omega_n^\Tr = \Omega_n^{n-1},
\end{align}
where for the rightmost equality we used the fact that a cyclic downwards-shift ($\Omega_n^\Tr x$) of a vector of length $n$ equals a cyclic upwards-shift by $n-1$ places ($\Omega_n^{n-1} x$). Using~\eqref{eq:jomega}, we obtain
\begin{equation*}
J\Omega_n^k J =
\Omega_n^{n-1} J\Omega_n^{k-1} J =\dots
= \Omega_n^{kn-k} =\left( \Omega_n^\Tr \right)^k
= \Omega_n^{n-k},
\end{equation*}
where we used $\Omega_n^{-1}=\Omega_n^\Tr$ and $\Omega_n^n=\I_n$. Consider a $\BC$ matrix $X\in\BC(n,l,m)$ represented as in~\eqref{eqa:bc2}. If $X$ is also to be $\CS$, then it must commute with the reflection matrix, i.e.
\begin{equation*}
\begin{aligned}
&\sum_{k=0}^{n-1}\Omega_n^k\otimes x_k \overset{!}{=}
J_{nl}\left(\sum_{k=0}^{n-1}\Omega_n^k\otimes x_k\right) J_{nm},\\
&=\left(J_{n}\otimes J_l\right)\left(\sum_{k=0}^{n-1}\Omega_n^k\otimes x_k\right) \left(J_{n}\otimes J_m\right),\\
&=\sum_{k=0}^{n-1}J_{n}\Omega_n^kJ_{n}\otimes J_{l}x_kJ_{m},\\
&= \sum_{k=0}^{n-1}\Omega_n^{n-k}\otimes J_{l}x_kJ_{m}=\sum_{k=0}^{n-1}\Omega_n^{n-k}\otimes J_{l}x_kJ_{m}.
\end{aligned}
\end{equation*}
Note that $\Omega_n^{j}$ has non-zero entries where $\Omega_n^{k}$, $k\neq j$, has zero entries and vice-versa. Equating the terms with the same power of $\Omega_n$ yields
\begin{align}\label{eq:bccsblocks}
x_{n-k} = J_l x_k J_m,
\end{align}
and in particular $x_0=J_l x_0 J_m$ and $x_{n/2}=J_l x_{n/2} J_m$.

\subsection{Decomposition of the Centrosymmetric Decomposition}\label{app:csbcdecomp}
When $\mathbf{R}$ is $\BC$ and $\CS$, the $\CS$ decomposition $\mathbf{\tilde{R}}=\trans{T}_{SN_B/2} \mathbf{R} T_{SN_C/2}$ can be further decomposed. Consider the partitioning of $\mathbf{R}$ into four equal-sized blocks as in Definition~\ref{def:cs} such that $\mathbf{\tilde{R}}=\diag\left(R_1-R_2J_{SN_C/2},R_1+R_2J_{SN_C/2}\right)$. From the $\BC$ structure~\eqref{eq:bc}, $R_1$ and $R_2$ are obtained as
\begin{align*}
R_1\!=\!
{\arraycolsep=2pt%
\left[
  \begin{array}{@{}*4c}
\mathbf{r}_0 & \mathbf{r}_1 & \dots & \mathbf{r}_{\frac{S}{2}\sm 1}\\
\mathbf{r}_{S\sm 1} & \mathbf{r}_0 & \dots & \mathbf{r}_{\frac{S}{2}-2}\\
\vdots & \vdots & \rotatebox[origin=c]{15}{$\ddots$} & \vdots\\
\mathbf{r}_{\frac{S}{2}+1} & \mathbf{r}_{\frac{S}{2}+2} & \dots & \mathbf{r}_{0}
  \end{array}
\right]},\,\,\,R_2\!=\!
{\arraycolsep=2pt%
\left[
  \begin{array}{@{}*4c}
\mathbf{r}_{\frac{S}{2}} & \mathbf{r}_{\frac{S}{2}+1} & \dots & \mathbf{r}_{S\sm 1}\\
\mathbf{r}_{\frac{S}{2}\sm 1} & \mathbf{r}_{\frac{S}{2}} & \dots & \mathbf{r}_{S-2}\\
\vdots & \vdots & \rotatebox[origin=c]{15}{$\ddots$} & \vdots\\
\mathbf{r}_{1} & \mathbf{r}_{2} & \dots & \mathbf{r}_{\frac{S}{2}}
  \end{array}
\right]}.
\end{align*}
The matrices $R_1,R_2$ have the $\CS$ blocks $\mathbf{r}_0$ and $\mathbf{r}_{S/2}$ on their diagonals. In addition, the blocks opposite the diagonals are $\mathbf{r}_k$ and $\mathbf{r}_{k-1}$ for $R_1$ and $\mathbf{r}_{S/2+k}$ and $\mathbf{r}_{S/2-k}$ for $R_2$, i.e. the opposite blocks satisfy the second part of Proposition~\ref{thm:reflblocks}. This entails that $R_1,R_2\in\CS(SN_B/4,SN_C/4)$. Note that if $R_2$ is $\CS$, then so is $R_2J_{S/2 N_C}$. Because the sum of two $\CS$ matrices is also $\CS$, the blocks of $\mathbf{\tilde{R}}$ are $\CS$ and each one of the blocks can be further decomposed by pre- and post-multiplication with $\trans{T}_{SN_B/4}$ and $T_{SN_C/4}$, respectively. In case $SN_B$ or $SN_C$ are not divisible by $4$, the decomposition is still possible but a different transformation matrix must be used (see for example~\cite{MIRRORMAT}).

\section{}\label{app:approx}
\subsection{Block-Circulant Approximation}\label{app:bcapprox}
For approximating a matrix $\mathbf{R}^p\in\R^{nl\times nm}$ with a matrix $X\in\mathcal{BC}(n,l,m)$, the optimization~\eqref{eq:ormapproxsec} is reformulated as
\begin{align}\label{eq:bcapprox}
\min_{\lbrace x_k\rbrace_{k=0}^{n\sm 1}}\frobnorm{\sum_{k=0}^{n-1}\Omega_n^k\otimes x_k - \mathbf{R}^p}^2,
\end{align}
where $\inR{x_k}{l}{m}$, $X$ was partitioned as in~\eqref{eq:bc} and the $\BC$ representation~\eqref{eqa:bc2} was used. Because $\Omega_n^k$ has non-zero elements where $\Omega_n^j$, $j\neq k$, has zero elements, i.e. $\sum_{k=0}^{S-1}\Omega_S^k = \mathbf{1}_{n,n}$, where $\mathbf{1}_{n,n}$ is a matrix of ones, problem~\eqref{eq:bcapprox} can be rewritten as
\begin{align}\label{eq:bcapproxsplit}
\min_{\lbrace x_k\rbrace_{k=0}^{n\sm 1}}
\sum_{k=0}^{n-1}\frobnorm{\Omega_n^k\otimes x_k -
\left(\Omega_n^k\otimes \mathbf{1}_{l,m}\right)\circ
\mathbf{R}^p}^2.
\end{align}
By partitioning $\mathbf{R}^p$ as
\begin{align}\label{bc:part}
\mathbf{R}^p =
\begin{bmatrix}
r_{0,0}       & r_{0,1}    & \dots  & r_{0,n-1}\\
r_{1,0}   & r_{1,1}    & \dots  & r_{1,n-1}\\
\vdots    & \vdots & \ddots & \vdots \\
r_{n-1,0}       & r_{n-1,1}    & \dots  & r_{n-1,n-1}\\
\end{bmatrix},
\end{align}
where $r_{i,j}\in\R^{l\times m}$, each summand on the right-hand side of~\eqref{eq:bcapproxsplit} can be rewritten as
\begin{align}\label{eq:bcapproxsplit2}
\sum_{j=0}^{n-1}
\frobnorm{x_k-r_{j,\,k+j\text{ mod } n}}^2,
\end{align}
for $k=0,\dots,n-1$. Using~\eqref{eq:bcapproxsplit}, \eqref{eq:bcapproxsplit2} and the partitioning~\eqref{bc:part}, the minimization~\eqref{eq:bcapprox} can be reformulated as
\begin{align}\label{eq:bcapproxsplit3}
\min_{\lbrace x_k\rbrace_{k=0}^{n\sm 1}}\sum_{k=0}^{n-1}\sum_{j=0}^{n-1}
\frobnorm{x_k-r_{k,\,k+j\text{ mod } n}}^2.
\end{align}
The minimum of~\eqref{eq:bcapproxsplit3} is attained where its derivative is zero. This can be done element-wise for each element of $x_k$ which -- after reconstructing blocks $x_k$ -- yields $x_k^\star = \frac{1}{n}\sum_{j=0}^{n-1}r_{k,\,k+j\text{ mod } n}$. Using the cyclic shift matrix, the solution is reconstructed as
$X^\star =
\sum_{k=0}^{n-1}
\left(\Omega_n^k\otimes\I_{l}\right)^\Tr
\mathbf{R}^p
\left(\Omega_n^k\otimes\I_{m}\right)/n$.

\subsection{Centrosymmetric Approximation}\label{app:csapprox}
For approximating a matrix $\mathbf{R}^p\in\R^{2q\times 2t}$ with a matrix $Y\in\mathcal{CS}(q,t)$, the optimization~\eqref{eq:ormapproxsec} is reformulated as
\begin{align}\label{eq:csapprox}
\min_{y_1,y_3\in\R^{q\times t}}\frobnorm{
\left[\begin{array}{c;{4pt/3pt}c}
y_{1} & J_q y_{3} J_t\\[0.3em]
\hdashline[4pt/3pt]\\[-1em]
y_{3} & J_q y_{1} J_t
\end{array}\right]
- \mathbf{R}^p}^2.
\end{align}
If $\mathbf{R}^p$ is partitioned as $\mathbf{R}^p =
\left[\begin{array}{c;{4pt/3pt}c}
r_{1} & r_2\\[0.3em]
\hdashline[4pt/3pt]\\[-1em]
r_{3} & r_4
\end{array}\right]$ with $r_i\in\R^{q\times t}$, the minimization~\eqref{eq:csapprox} can be reformulated as
\begin{equation}\label{eq:csapproxsplit}\begin{aligned}
\min_{y_1,y_3\in\R^{q\times t}} &\left(
\frobnorm{y_1-r_1}^2 +
\frobnorm{y_{3}-J_q r_2 J_t}^2 \right.\\
&\left.\quad+\frobnorm{y_3-r_3}^2 +
\frobnorm{y_{1} - J_q r_4 J_t}^2
\right)
\end{aligned}\end{equation}
where $\frobnorm{UXV}=\frobnorm{X}$ for orthonormal $U,V$ was used~\cite[Ch. 2.3.5, p.~75]{GOLUB4}. The minimum of~\eqref{eq:csapproxsplit} is attained where its derivative is zero. The minimizers $y_1^\star$ and $y_2^\star$ are obtained as $y_1^\star =\left(r_1 + J_q r_4 J_t\right)/2$ and $y_3^\star =\left(r_3 + J_q r_2 J_t\right)/2$ and $Y^\star$ is reconstructed as $Y^\star = \left(\mathbf{R}^p + J_{2q} \mathbf{R}^p J_{2t}\right)/2$.

\subsection{Block-Circulant and Centrosymmetric Approximation}\label{app:bccsapprox}
For approximating a matrix $\mathbf{R}^p\in\R^{nl\times nm}$, where $n,l,m>1$ are even, with a matrix $Z\in\BC(n,l,m)\cap\CS(nl/2,nm/2)$, the optimization~\eqref{eq:ormapproxsec} is reformulated as in~\eqref{eq:bcapprox} and the blocks $\lbrace z_k\rbrace_{k=n/2+1}^{n\sm 1}$ are substituted using~\eqref{eq:bccsblocks}, which yields
\begin{equation}\label{eq:bccsapprox}
\begin{aligned}
&\min_{\lbrace x_k\rbrace_{k=0}^{n\sm 1}}\frobnorm{\I_n\otimes z_0+\Omega_n^{n/2}\otimes z_{n/2}\\
&\quad+\sum_{k=1}^{n/2-1}\left(\Omega_n^k\otimes z_k +\Omega_n^{n-k}\otimes J_l z_k J_m\right)- \mathbf{R}^p}^2,\\
&=\min_{\lbrace x_k\rbrace_{k=0}^{n\sm 1}}\Bigg(
\frobnorm{\I_n\otimes z_0 - \left(\I_n\otimes\mathbf{1}_{l,m}\right)\circ\mathbf{R}^p}^2\\
&\quad+\frobnorm{\Omega_n^{n/2}\otimes z_{n/2} - \left(\Omega_n^{n/2}\otimes\mathbf{1}_{l,m}\right)\circ\mathbf{R}^p}^2\\
&\quad+\sum_{k=1}^{n/2-1}\bigg(
\frobnorm{\Omega_n^k\otimes z_k - \left(\Omega_n^k\otimes\mathbf{1}_{l,m}\right)\circ\mathbf{R}^p}^2\\
&\quad+\frobnorm{\Omega_n^{n-k}\otimes J_l z_{k} J_m - \left(\Omega_n^{n-k}\!\otimes\!\mathbf{1}_{l,m}\right)\circ\mathbf{R}^p}^2
\bigg)\!\Bigg).
\end{aligned}
\end{equation}
As for the $\BC$ approximation in Appendix~\ref{app:bcapprox}, the Frobenius norms can be separated for different powers of $\Omega_n$. The terms for $z_0$ and $z_{n/2}$ can be rewritten as
\begin{align}\label{eq:bccsz0}
\frobnorm{\Omega_n^{k}\otimes z_{k} - \left(\Omega_n^{k}\otimes\mathbf{1}\right)\circ\mathbf{R}^p}^2 =
\sum_{j=0}^{n-1}\frobnorm{z_k-\rho_{k,j}}^2,
\end{align}
where $k=\lbrace 0,n/2\rbrace$, $\rho_{k,j}\eqdef r_{k,\,k+j\text{ mod } n}$ with $\mathbf{R}^p$ partitioned as in~\eqref{bc:part}. According to~\eqref{eq:bccsblocks}, sub-blocks $z_0$ and $z_{n/2}$ must be $\CS$. Sub-blocks $z_k$ and $\rho_{k,j}$ are therefore partitioned as
\begin{align*}
z_k =
\left[\begin{array}{c;{4pt/3pt}c}
z_k^1 & J_{l/2} z_k^3 J_{m/2}\\[0.3em]
\hdashline[4pt/3pt]\\[-1em]
z_k^3 & J_{l/2} z_k^1 J_{m/2}
\end{array}\right],\quad
\rho_{k,j} =
\left[\begin{array}{c;{4pt/3pt}c}
\rho_{k,j}^1 & \rho_{k,j}^2\\[0.3em]
\hdashline[4pt/3pt]\\[-1em]
\rho_{k,j}^3 & \rho_{k,j}^4
\end{array}\right],
\end{align*}
and the right-hand side of~\eqref{eq:bccsz0} rewritten as
\begin{equation}\label{eq:z0finalsum}\begin{aligned}
\sum_{j=0}^{n-1}
\big(
&\frobnorm{z_k^1-\rho_{k,j}^1}^2 +
\frobnorm{z_k^3-J_{l/2} \rho_{k,j}^2 J_{m/2}}^2\\
&+\frobnorm{z_k^3-\rho_{k,j}^3}^2 +
\frobnorm{z_k^1 - J_{l/2} \rho_{k,j}^4 J_{m/2}}^2
\big).
\end{aligned}\end{equation}
Note the similarity between~\eqref{eq:z0finalsum} and~\eqref{eq:csapproxsplit}. Setting the derivative of~\eqref{eq:z0finalsum} to zero, solving for $z_k^{1},z_k^{3}$ and reconstructing $z_k$ yields for $k=\lbrace 0,n/2\rbrace$
\begin{align}
z_k^\star = \frac{1}{2n}\sum_{j=0}^{n-1} \left(\rho_{k,j} + J_l \rho_{k,j} J_m \right).\label{eq:z0approxresult}
\end{align}
The summands in~\eqref{eq:bccsapprox} for $k\!=<!1,\dots,n/2\sm 1$ are rewritten as
\begin{align*}
&\sum_{j=0}^{n-1}\frobnorm{z_k - \rho_{k,j}}^2+
\underbrace{\frobnorm{J_l z_{k} J_m - \rho_{n-k,j}}^2}_{=\frobnorm{z_{k} - J_l \rho_{n-k,j} J_m}^2}.
\end{align*}
Setting the derivative to zero yields for $k=1,\dots,n/2\sm 1$
\begin{align*}
z_k^\star = \frac{1}{2n}\sum_{j=0}^{n-1} \left(\rho_{k,j} + J_l \rho_{n-k,j} J_m \right),
\end{align*}
which is identical to~\eqref{eq:z0approxresult}. After reconstruction, the matrix $Z^\star$ is obtained as
\begin{align*}
Z^\star =
\frac{1}{2n}
\sum_{k=0}^{n-1}
\left(\Omega_n^k\otimes\I_{l}\right)^\Tr
\left(
\mathbf{R}^p +
J_{nl}
\mathbf{R}^p
J_{nm}
\right)
\left(\Omega_n^k\otimes\I_{m}\right).
\end{align*}

\end{document}